\documentclass[aps,prl,twocolumn,showpacs]{revtex4}
\usepackage{color}
\usepackage{graphicx}
\usepackage{bm}

\begin{document}


%
\title{Why metallic surfaces with grooves a few nanometers deep and wide may strongly absorb visible light.}
\author{J. Le Perchec,  P. Qu\'emerais, A. Barbara, T. L\'{o}pez-R\'{\i}os}
\address{Institut N\'eel, CNRS and Universit\'e Joseph Fourier, BP 166, 38042 Grenoble Cedex 9, France}
%

\begin{abstract}It is theoretically shown that nanometric silver lamellar gratings
present very strong visible light absorption inside the grooves,
leading to electric field intensities several orders of magnitude
larger than that of the impinging light. This effect, due to the
excitation of quasi-static surface plasmon polaritons with
particular small penetration depth in the metal, may explain the
abnormal optical absorption observed a long time ago on almost flat
Ag films. Surface enhanced Raman scattering in rough metallic films
could also be due to the excitation of such quasi-static plasmon polaritons in
grain boundaries or notches of the films.
\end{abstract}
\maketitle In general, modifications of metallic surfaces at
nanometer scales lead to negligible changes in the reflectivity of
the visible and infrared light. However, when the impinging light is
combined to surface electromagnetic modes to give rise to surface
plasmons-polaritons (SPPs), the optical response can become very
surface sensitive. SPPs arouses a lot of interest as they could play
a key role in the issue of merging optics to
electronics\cite{ozbay}. The particular case of long wave vectors
has been recently investigated and interesting theoretical and
experimental works devoted to electromagnetic plane wave guides with
nanometer dimensions\cite{tanaka,pile,miyazaki1,maier,miyazaki2} or
in the microwave regim\cite{sambles} were published. The underlying
physics of these highly sub-wavelength guides is that of the coupled
SSPs taking birth at the two dielectric/metal interfaces of a
metal-insulator-metal system. This coupling splits the dispersion
curve of the unique SPP into a symmetric and an anti-symmetric
branch\cite{economou,prade}. For sub-wavelength thicknesses of the
insulator, the unique guided mode is built by the anti-symmetric
branch whose dispersion shifts towards the long wave vector as the
thickness decreases. In a different context, SPPs were also
considered in an attempt to understand the surface enhanced Raman
scattering (SERS) and it is currently admitted they are involved in
its basic mechanism. SPPs should also be related to a much older
misunderstood phenomena: the abnormal optical absorption (AOA) of
alkali metals deposited on a cold glass wall which present
absorption bands independently of the incidence angle which can not
be attributed to diffraction or any another known effect\cite{wood}.
More recently, this abnormal absorption was observed for other
metals\cite{myers}. Silver films presenting AOA and SERS made by
vapour quenching on cold substrates show a typical roughnesses of
shape of 5-30 nm when observed in situ with a STM\cite{douketis}.
This is one of the numerous indications that SERS may occur for
molecules adsorbed on surfaces presenting a very small amplitude
roughness\cite{moskovits}. Up to now, these observations remained
partly mysterious because of the nanometer size of the geometrical
shapes involved in these phenomena \cite{moskovits}. The absorption
of light by SPPs propagating on a flat metallic surface, using a
prism, is known since a long time \cite{otto}. Later, following the
pioneering work of Hessel and Oliner \cite{hessel}, Wirgin et al.
\cite{wirgin} showed that cavity (Fabry-Perot like) modes excited
inside grooves made on a metallic surface may also participate to
the visible light absorption. This was confirmed by next studies
\cite{sobnack, tan, astilean, hooper,popov}. However, in all these works,
either the grooves depth $h$ was about 100-400 nm, and roughly related to
the exciting light wavelength by $h \sim\lambda/4$, which is the
usual Fabry-Perot resonance condition \cite{wirgin} for these kinds
of cavities, either excitation of SPPs propagating on the upper horizontal surface of the gratings was considered.
\\In the present
paper, we show for the first time that cavities only a few nanometers deep ($\sim 5-15$ nm) and wide ($\sim 2-5$ nm), i.e. with depth one order smaller than those considered in all previous studies, may also act as guides and resonators leading to a very strong absorption of visible
light ($\lambda \sim 500$ nm). Electric field intensities in excess of thousands times
larger than that of the impinging light may exist and is confined inside the cavities. We show that it is due to the excitation of SPPs in the electrostatic (quasi-static) regime .
\\Our results were obtained by the exact modal method, originally developed by Botten et al. \cite{botten} and
Sheng et al. \cite{sheng}, for the grating depicted in fig.1,
submitted to a p-polarized wave (electric field perpendicular to the
grooves). Space is divided into three regions: above ($y>0$) and
below ($y<-h$) the grating (regions I and III respectively) wherein
the magnetic field $H_z$ is expressed as a Rayleigh plane wave
expansion, and region II of the grating ($-h<y<0$) wherein $H_z$ is
expressed by a modal expansion. The electric field is obtained from $H_z$ by means of Maxwell's equations.
\begin{figure}
\centering\includegraphics[scale=0.22]{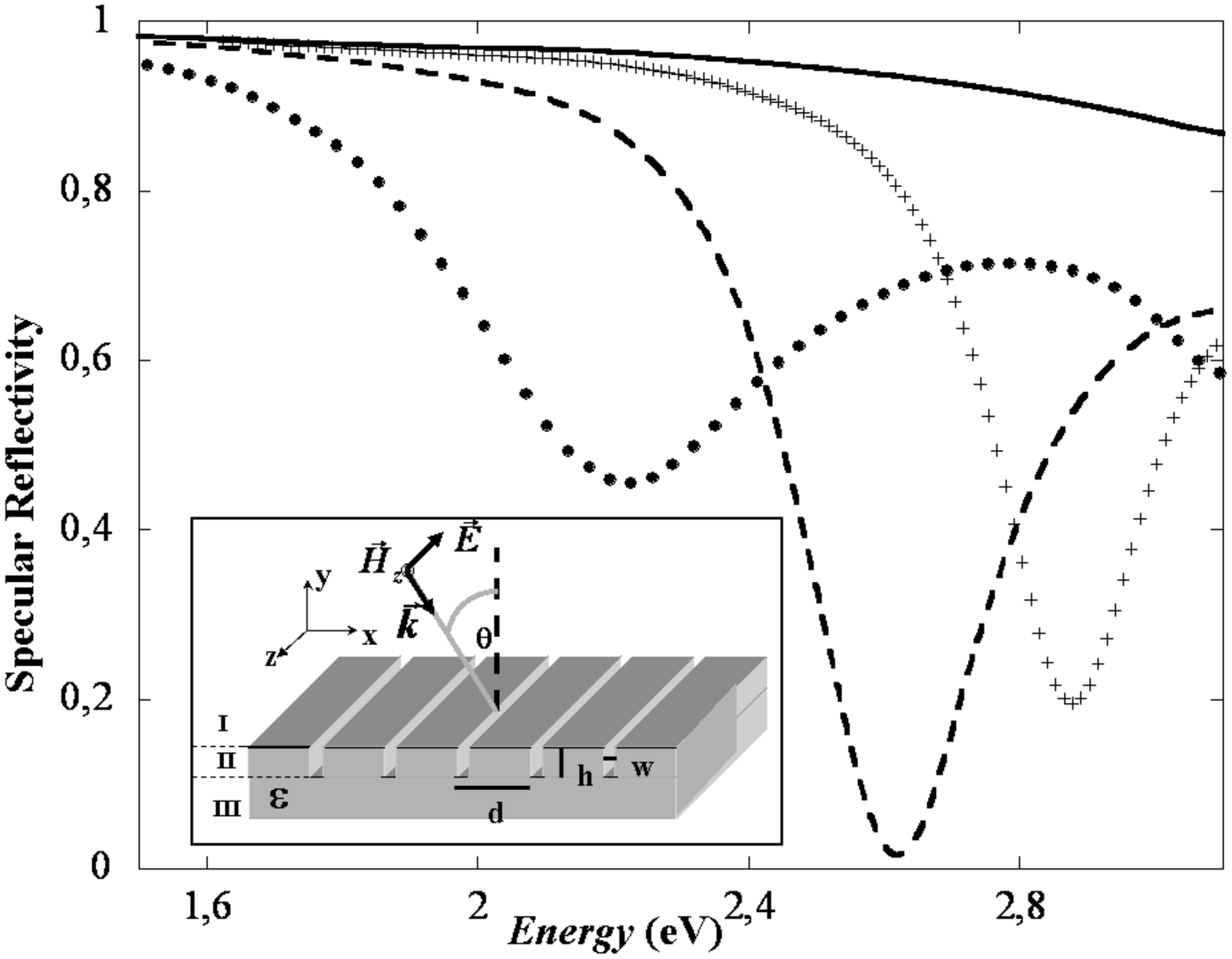}
\caption{Reflectivity of a p-polarized plane wave impinging on the
silver grating, schematic represented in the insert, at normal
incidence calculated with an exact model for different heights
$h=30$ nm ($\bullet$), $h=15$ nm ($---$) and $h=5$ nm ($+$) with
$w=5$ nm and $d=30$ nm. The black curve corresponds to the
reflectivity of a flat silver film.}
\end{figure}
With $\epsilon_{air}=1$ for sake of simplicity
and $\epsilon_{metal}=\epsilon$, the fields are given by
\cite{sheng}:
\begin{eqnarray}
H_z^I(x,y)&=&e^{ik(\gamma_0x-\eta_0^Iy)}+
\sum_{n=-\infty}^{+\infty}R_ne^{ik(\gamma_nx+\eta_n^Iy)} \nonumber \\
H_z^{III}(x,y)&=&\sum_{n=-\infty}^{+\infty}T_ne^{ik(\gamma_nx-\eta_n^{III}(y+h))} \nonumber \\
H_z^{II}(x,y)&=&\sum_{\ell=0}^{+\infty}X_{\ell}(x)\left(A_{\ell}e^{i
\Lambda_{\ell}y}+B_{\ell}e^{-i\Lambda_{\ell}y}\right),\nonumber
\end{eqnarray}
where $\gamma_n= \sin{\theta} + 2 \pi n/d$, $\eta_n^{I}= \sqrt{1-
\gamma_n^2}$ and $\eta_n^{III}=\sqrt{\epsilon-\gamma_n^2}$.
$H_z^{II}$ is a linear combination of the eigenmodes $\{X_{\ell}\}$, each of them being characterized by its
eigenwave-vector $k_y=\Lambda_{\ell}$. These are solutions of the
eigenvalue equation \cite{botten,sheng}:
\begin{eqnarray}
\cos(kd \sin\theta)&=&\cos(\beta_{\ell}(d-w))\times
\cos(\alpha_{\ell} w)
\\&-&\frac{1}{2}\left[\frac{\alpha_{\ell}\epsilon}{\beta_{\ell}}+\frac{\beta_{\ell}}{\alpha_{\ell}\epsilon}\right]\times \sin(\beta_{\ell}(d-w)) \sin(\alpha_{\ell} w) \nonumber
\end{eqnarray}
where $\alpha_{\ell}^2=k^2-\Lambda_{\ell}^2$, and
$\beta_{\ell}^2=k^2 \epsilon- \Lambda_{\ell}^2$ and $\theta$ is the
incidence angle. Once the fields are expressed in the three regions,
we employ the boundary conditions at the horizontal interfaces and
project each resulting equation on two different basis \cite{theseJerome}. That
allows to determine unambiguously the coefficients
$\{A_{\ell}\}$, $\{B_{\ell}\}$, $\{ R_n \}$ and $\{T_n \}$, for
$\ell \in [0,L]$ and $n \in [-N,+N]$. Convergency of the
solution has been checked by increasing $L$ and $N$.
Typically, $N \sim 400$, and only few modes $L \sim 30$ are
enough for all considered cases. For very
small $w$, only the fundamental mode $\ell=0$ plays a
significant role whereas all the others $\ell>0$ are
strongly evanescent in the grooves. We have tested the
method by comparing our numerical results with those obtained by two
other accurate numerical methods (\textit{RCWA} and \textit{FDTD})
for metallic gratings \cite{astilean, vanlabecke}. Fig 1 shows the
calculated reflectivity at normal incidence of a silver grating with
$d= 30$ nm, $w =5$ nm and $h= 30$, $15$ and $5$ nm. In all cases, we
choose $d<<\lambda$ such that SPPs at the horizontal interface $y=0$
are never excited in the range of the considered frequencies. The
figure shows that noticeably amount of photons can be absorbed by
this very small amplitude grating at specific wavelength in the
visible spectrum. It is to note that for $h=15$ nm the reflectivity
is almost zero at $\sim 2.6$ eV ($480$ nm) whereas that
of the flat silver plane stays close to 1. This is a reliable
quantitative result for AOA.
\begin{figure}
\centering\includegraphics[scale=0.25]{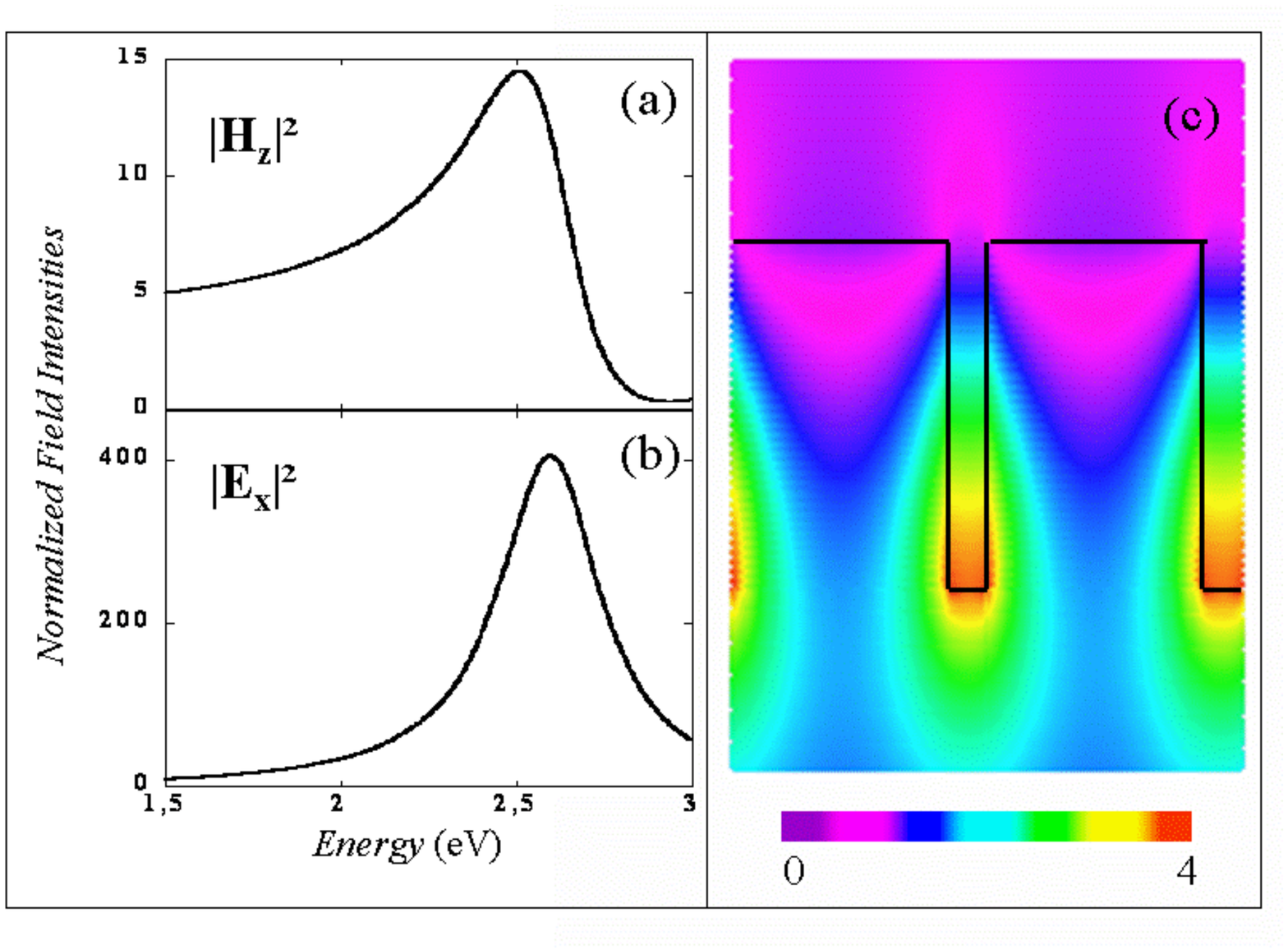}
\caption{Normalized intensities with respect to the incident field
of the magnetic field at $y=-h$ (a) and of the electric field along
the x-axis at $y=0$ (b), calculated for the grating $h=15$ nm, $w=5$
nm and $d=30$ nm. (c) Map of the normalized magnetic field modulus
near the grating, at the resonance ($\omega=2.6$ eV).}
\end{figure}
Fig.2 illustrates that the absorption is due to resonances within
the tiny Ag grooves. Indeed, the reflectivity falls are associated
to enhancements of the magnetic and electric fields intensity inside
the grooves. Enhancements of more than two orders of magnitude are
achieved for the electric field while the magnetic field is only
enhanced by a factor $10-20$. The spatial distribution of the
normalized magnetic field modulus is represented fig.2c considering
the grating with $h=15$ nm and at the resonant energy $\omega=2.6$
eV. One clearly sees that the very sub-wavelength cavities resonate
and absorb a great part of the incident photons.
\\In order to understand why such strong resonances may occur for such
shallow grooves, let us return to the SPP dispersion of a perfectly
flat metal/vacuum interface. For simplicity we take the dielectric
constant of the metal negative and real ($- \infty<\epsilon<-1$).
The dispersion is given by the explicit well-known relation
$k_{//}=k \sqrt {\epsilon/ (\epsilon+1)}$, where $k_{//}$ is the SPP
wave vector parallel to the interface. As it is
known \cite{economou}, we may distinguish two asymptotic regimes: the
"optical regime" when $\epsilon \rightarrow -\infty$, and the
"electrostatic regime" when $\epsilon \rightarrow -1$. In the
optical regime, retardation effects play a significant
role\cite{economou}. The electromagnetic fields at the interface
satisfy $\vert E_{\bot}
 /H \vert = k_{//} \approx \omega/c$ and the excited plasmon
has a structure very similar to that of light. On the opposite side
in the electrostatic limit, retardation effects remain
negligible \cite{economou}. We have $\vert E_{\bot}/ H \vert
\rightarrow \infty$ and the obtained plasmons have essentially an
electric component. In this context, a relevant physical quantity to
introduce is the dimensionless ratio $X= \delta_p / \delta_s$, where
$\delta_p$ is the penetration depth of the SPP in the metal, and
$\delta_s$ is the usual skin depth in the metal of a plane wave
whose wave number is $k$: $\delta_s^2 = 1/ \vert \epsilon \vert k^2$
while $\delta_p^2=-1/ k_{\bot}^2$. For a flat interface, $X=\sqrt
{(\epsilon +1)/ \epsilon}$. The parameter $X$, satisfies $0<X<1$,
and completely determines the two regimes: in the optical regime, $X
\rightarrow 1$ and $\delta_p \approx \delta_s$ while in the
electrostatic regime $X \rightarrow 0$ and $\delta_p \ll \delta_s$.
\begin{figure}
\centering\includegraphics[scale=0.25]{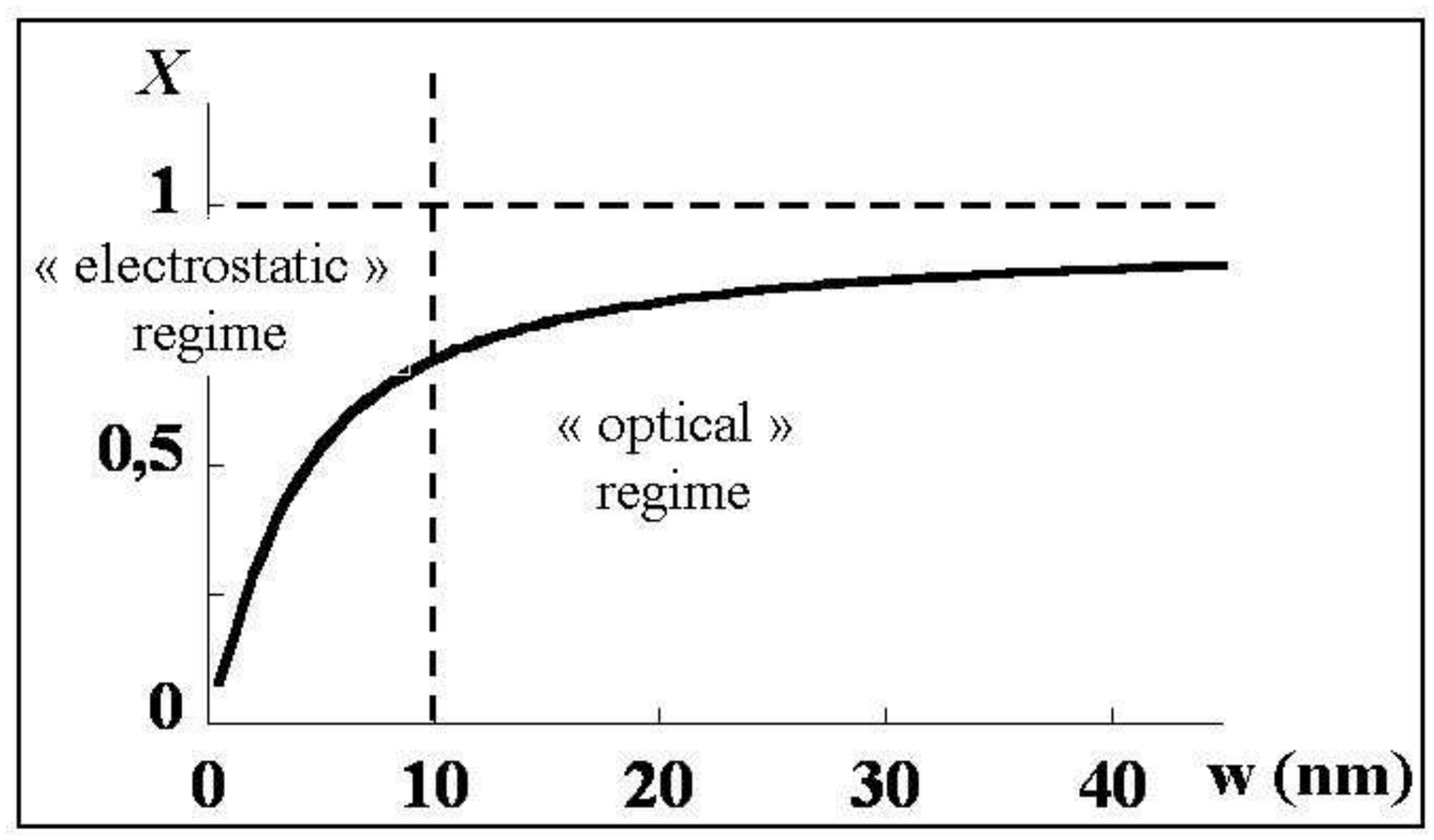}
\caption{Evolution of the dimensionless parameter
$X=\delta_p/\delta_s$ as a function of the width $w$ of the grooves
calculated for $\omega=2.48$ eV. Modifying $w$ allows to move
continuously from one regime to the other, at a \textit{given}
frequency.}
\end{figure}
\\Returning to our grooves with small $w$, we show that the role
played by the decrease of $w$ down to the nanometer scale, is to
displace the dispersion of the mode guided in the cavities, from the
optical regime to the electrostatic one, for a given frequency and
thus for a \textit{fixed} $\epsilon$ value (figure 3). Let us
consider the dispersion relation Eq.(1) of the grating in the case
of silver and in a range of wave numbers $k$ for which values of
$\epsilon$ are typically in the interval $-25<\epsilon<-5$ (visible
range). Here again we neglect the imaginary part of $\epsilon$, we
will come back to this approximation later. In the case of
sub-wavelength values of $w$, it is easy to show that
$\Lambda_{\ell>0}^2<0$ and that the only guided wave in the groove
is the fundamental mode whose wave vector is $\Lambda_0^2>0$. This
mode may satisfy a Fabry-Perot resonance condition when $\Lambda_0 h
\sim \pi/2$, as for gratings with deep
grooves\cite{wirgin,astilean}, leading to the cavity
resonance we are discussing. The problem is to determine the value
of $\Lambda_0$ and its dispersion. For large enough $d$ and at
normal incidence, Eq.(1) may be simplified and factorized, so that
the fundamental mode fulfills:
\begin{displaymath}
\tan {\left( \frac{\alpha_0w}{2} \right)}+\frac{i \beta_0}{\epsilon
\alpha_0} \approx 0,
\end{displaymath}
where $\alpha_0^2=k^2-\Lambda_0^2$, $\beta_0^2 = \epsilon k^2-
\Lambda_0^2$. For small values of $\vert \alpha_0 \vert w$ (which is
always verified for sub-wavelength $w$), we get a simple second
degree equation for $\beta_0$ : $\beta_0^2+(2i/\epsilon w)
\beta_0-(\epsilon-1) k^2 \approx 0$. We now introduce the same
quantity $X$ as for the perfectly flat plane: $X= \delta_p/\delta_s=
1/ \vert \beta_0 \vert \delta_s$. Solving the previous second degree
equation in $X$ we obtain:
\begin{equation} X= \frac
{\delta_p}{\delta_s} = \sqrt{ \frac{ \epsilon}{\epsilon-1}}
f(\Gamma),
\end{equation}
where $f(\Gamma)= -\Gamma+ \sqrt{ \Gamma^2+1}$ and $\Gamma =
\delta_s / \sqrt{ \epsilon (\epsilon-1)}w$. Notice that Eq.(2) is an
almost exact result which is valid whatever $k$ and $\epsilon$ are,
provided that the imaginary part of $\epsilon$ remains small with
respect to its real part, and that $w<\lambda$. The function
$f(\Gamma)$ satisfies $0<f(\Gamma)<1$ and fully characterizes the
behavior of the system. The dimensionless parameter $\Gamma$
reflects the strength of the coupling between the SPPs propagating
on the two vertical walls of the grooves. For $\Gamma\ll 1$, the
coupling is weak, whereas at strong coupling $\Gamma\gg 1$.
\begin{figure}
\centering\includegraphics[scale=0.25]{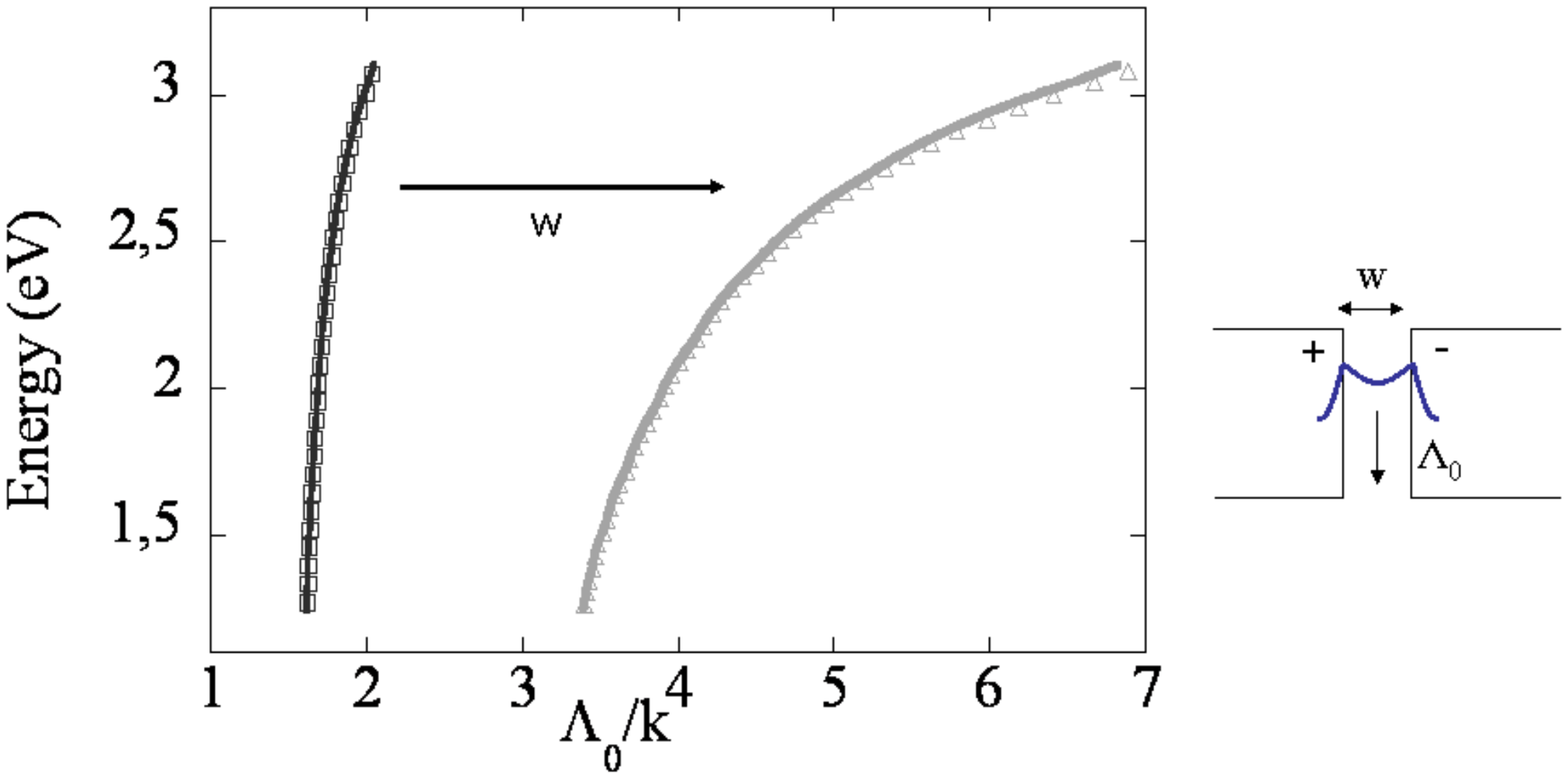}
\caption{Dispersion curve of the fundamental guide mode in the
grooves calculated for a period $d=300$ nm and two different widths:
$w=30$ nm (black curve) and $w=5$ nm (gray curve). Dots correspond
to the analytical calculation and lines correspond to the exact
numerical one.}
\end{figure}
Figure 3 depicts the behaviour of $X$ as function of $w$ for silver
at $1/\lambda=20000$ cm$^{-1}$ ($\omega=2.48$ eV, $\epsilon=-8.57$).
By increasing the coupling of the SPPs via the reduction of the
widths of the cavities, we can fully scan the different behaviour of
the usual SPP, from the electrostatic regime to the optical one,
with a crossover located around $w=10$ nm, and that at a given
frequency. This allows to have a control of the absorption
properties of the grating by a simple choice of geometrical
parameters.
\\Figure 4 depicts the dispersion of the
fundamental mode, which is analytically given by:
\begin{equation}
\Lambda_0 = \sqrt{\epsilon k^2-\beta_0^2}= \frac{1}{\delta_s} \sqrt{
\frac{1}{X^2}-1}.
\end{equation}
as well as the dispersion obtained by the numerical calculation, for
two different values of $w$ ($w=5$ and $w=30$ nm), and for $d=300$
nm. The agreement between the numerical and analytical curves is
excellent and shows that as $w$ decreases the wave vector of the
guided mode $\Lambda_0$ increases even though it is excited by the
same incident energy. Meanwhile, we know from fig.3 that its
penetration depth in the metal becomes much smaller than the
ordinary skin depth ($X\ll1$). The dispersion relation of the modes
in the optical regime can be deduced from Eq. 3 taking $\Gamma\ll 1$
and was already discussed\cite{astilean,sobnack, collin}. Reversely, in the electrostatic
regime, $\Gamma
>>1$, $f(\Gamma) \approx 1/2 \Gamma$ and Eq.(3) leads to:
\begin{equation}
\Lambda_0 \approx \frac{2}{\vert \epsilon\vert w},
\end{equation}
which implies that Fabry-Perot resonances
occur for very small values of $h$, correspondingly to those
obtained numerically on Fig.1 and Fig.2. The electromagnetic field
in the groove is dominated by the electric field, $\vert E_x /H
\vert\gg k$. It is also interesting to notice that since $\Lambda_0$
essentially depends on $w$, we may obtain a scaling law using the
condition resonance $\Lambda_0h \sim \pi/2$. All grooves with nearly
the same ratio $h/w$ will resonate around the same frequency. This
is shown on figure 5, for a grating with $h=18$ nm and $w=5$ nm and
one with $h=3$ nm and $w=1$ nm.
\begin{figure} \centering\includegraphics[scale=0.25]{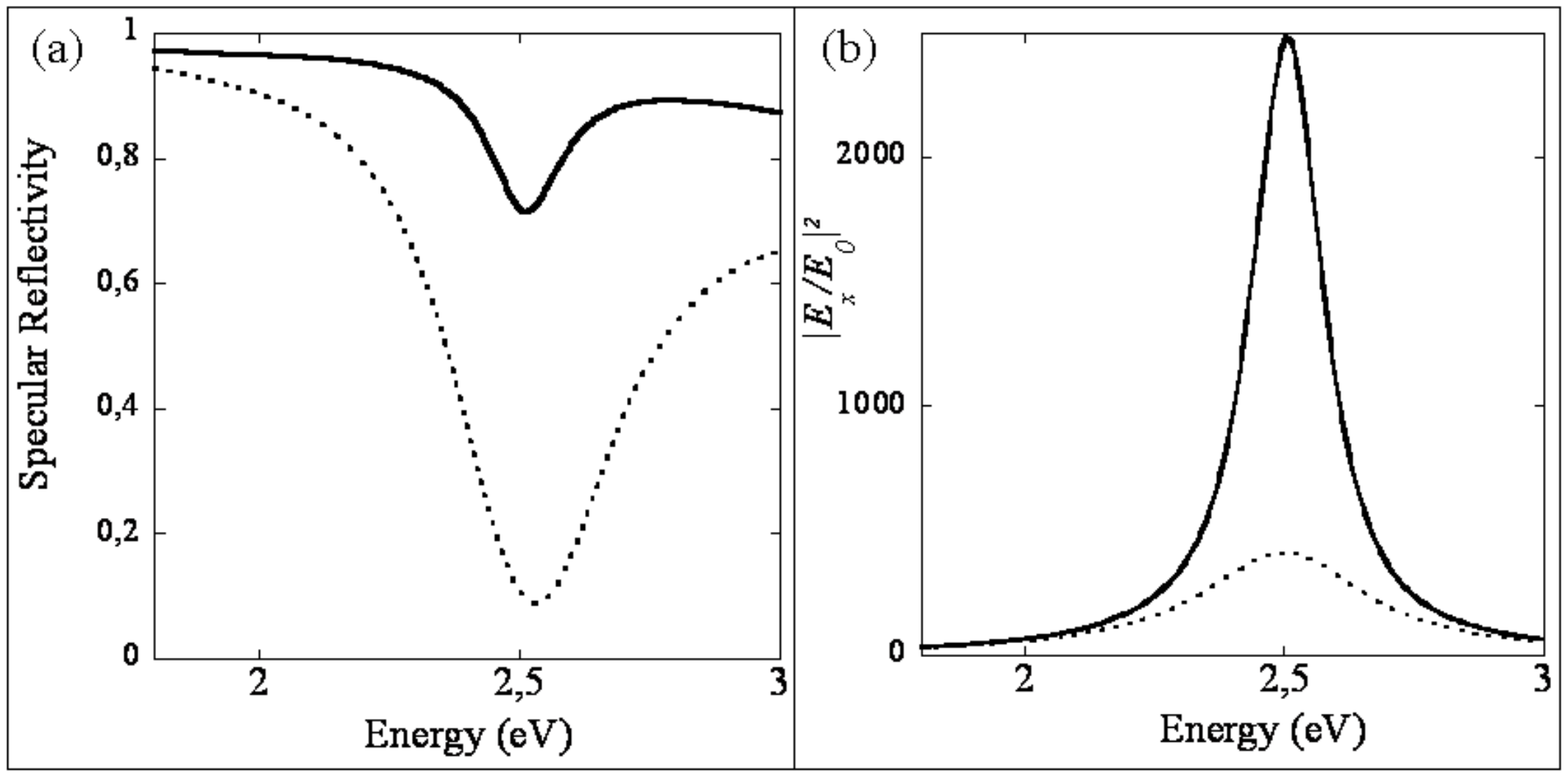}
\caption{(a) Reflectivity at normal incidence of two different
gratings with nearly the same $h/w$ ratio. $h=18$ nm and $w=5$ nm
(dashed line) and $h=3$ nm and $w=1$ nm (full line) both resonating
at the same frequency. (b) normalized intensity of the electric
field along the x-axis at the mouth of the cavities for the two same
gratings. Smaller cavities localize much stronger fields (imaginary part of $\epsilon$ is taken into account).}
\end{figure}.\\An analytical study of the fields expression
shows that $\vert E_x /E_{inc.} \vert \sim 2d/w$, where $E_{inc.}$
is the amplitude of the total incident field. Considerable electric
field enhancements can thus be obtained inside the grooves with small
$w$. Actually the electrostatic regime is easily obtained provided
that the sub-wavelength cavities are weakly coupled through the
metal, that is to say if the grooves are sufficiently distant
($\delta_s<d<\lambda$), otherwise $\Lambda_0$ stays around $k$.
Giant enhancements, obtained for large $d/w$, are numerically
observed: for instance, it is much greater than $10^4$ for $w=1$ nm,
$h=8$ nm and $d=200$ nm at $\omega=1.85$ eV ($\lambda\sim 670$ nm).
Finally, it should be observed that turning on a small imaginary
part of $\epsilon= \epsilon'+i \epsilon''$, with $\epsilon'' \ll
\vert \epsilon' \vert$, the resonant wavevector $\Lambda_0$ also has
an imaginary part $\Lambda''_0$. We can show analytically that for
$\Gamma>>1$: $ \Lambda_0''/ k''_{plane} \sim 4/kw $, where
$k''_{plane}= k \epsilon''/ (\epsilon')^2$ is the imaginary part of
the SPP wave vector of a perfectly flat surface. The attenuation of
the SPPs along the walls of the cavity is thus more important than
in the case of a single plane surface, for a given frequency.
Nevertheless the depth of our channels is as small as the wave
vector is long to excite the Fabry-Perot like resonance at
$\Lambda_0h\sim\pi/2$. Consequently these modes remain slightly
attenuated over the distance corresponding to the depth of the
channel.\\In conclusion, free electron metal surfaces with grooves
of rectangular shape and nanometer dimensions may absorb visible
light of  well defined frequencies and lead to extremely high electromagnetic
near-fields. Our calculations suggest that AOA observed on rather
smooth metal films may be due to notches (distorted grain
boundaries) few nanometers deep only. We have pointed out
some geometries that could optimize the near-field to generate
controlled Raman scattering enhancements. Finally, the calculated
enhancements suggest that SERS could also be due to the excitation
quasi-static surface polaritons in the grooves (giving rise to the
so-called "hot spots"), with
penetration depth much smaller than the usual skin depth.
\\The authors wish to thank A. Wirgin, Ph. Lalanne and S. Collin for
useful discussions.

\end{document}